\documentclass[a4paper,twoside]{article}

\usepackage{epsfig}
\usepackage{subcaption}
\usepackage{calc}
\usepackage{amssymb}
\usepackage{amstext}
\usepackage{amsmath}
\usepackage{amsthm}
\usepackage{multicol}
\usepackage{pslatex}
\usepackage{url} 
\usepackage{hyperref}
\usepackage{epstopdf}
\usepackage{apalike}
\usepackage{algorithm2e}
\usepackage[bottom]{footmisc}
\usepackage{SCITEPRESS}     

\begin{document}

\title{AI Agents with Decentralized Identifiers and Verifiable Credentials}

\author{\authorname{Sandro Rodriguez Garzon\sup{1}\orcidAuthor{0000-0001-6921-294X}, Awid Vaziry\sup{1}\orcidAuthor{0009-0007-2192-5968}, Enis Mert Kuzu\sup{2}, Dennis Enrique Gehrmann\sup{2}, Buse Varkan\sup{2}, Alexander Gaballa\sup{2} and Axel Küpper\sup{1}\orcidAuthor{0000-0002-4356-5613}}
\affiliation{\sup{1}Service-centric Networking / T-Labs, Technische Universität Berlin, Germany}
\affiliation{\sup{2}Technische Universität Berlin, Germany}
\email{\{sandro.rodriguezgarzon, vaziry, axel.kuepper\}@tu-berlin.de, \\\{enis.mert.kuzu, gehrmann, buse.varkan, gaballa\}@campus.tu-berlin.de}
}


\abstract{A fundamental limitation of current LLM-based AI agents is their inability to build differentiated trust among each other at the onset of an agent-to-agent dialogue. However, autonomous and interoperable trust establishment becomes essential once agents start to operate beyond isolated environments and engage in dialogues across individual or organizational boundaries. A promising way to fill this gap in Agentic AI is to equip agents with long-lived digital identities and introduce tamper-proof and flexible identity-bound attestations of agents, provisioned by commonly trusted third parties and designed for cross-domain verifiability. This article presents a conceptual framework and a prototypical multi-agent system, where each agent is endowed with a self-sovereign digital identity. It combines a unique and ledger-anchored W3C Decentralized Identifier (DID) of an agent with a set of third-party issued W3C Verifiable Credentials (VCs). This enables agents at the start of a dialog to prove ownership of their self-controlled DIDs for authentication purposes and to establish various cross-domain trust relationships through the spontaneous exchange of their self-hosted DID-bound VCs. A comprehensive evaluation of the prototypical implementation demonstrates technical feasibility but also reveals limitations once an agent’s LLM is in sole charge to control the respective security procedures.}  

\onecolumn \maketitle \normalsize \setcounter{footnote}{0} \vfill

\section{\uppercase{Introduction}}
\label{sec:introduction}

The advent of LLM-based AI agents has opened new opportunities for enhancing human productivity and decision-making \cite{10.1093/qje/qjae044}. A single agent can autonomously retrieve information, perform calculations, automate workflows, and interface with digital services on behalf of its user. Such capabilities already demonstrate tangible benefits in personal assistance and task execution. However, the true potential of agents emerges when they are able to interact and collaborate among each other across organizational boundaries \cite{collaborativeAgenticAI2025}. Through agent-to-agent communication, tasks that exceed the scope of an individual agent can be accomplished collaboratively. For instance, while one agent may support a user in planning a vacation by identifying preferences and budget constraints, it is only through secure and trusted interaction with external agents of travel agencies, airlines, and hotels that the full process of planning and booking a holiday can be autonomously and firmly executed.

This shift from isolated single-agent operation to cross-domain multi-agent collaboration comes with new challenges such as interoperability \cite{collaborativeAgenticAI2025}, security \cite{ko2025seven} \cite{kong2025survey} \cite{schroederdewitt2025open}, and trust \cite{huang2025zero} \cite{raza2025trism}. An important step toward addressing interoperability is the recently published Agent2Agent (A2A) protocol\footnote{\url{https://a2a-protocol.org}}, which provides a standardized interaction layer among agents that remains agnostic to the underlying agent frameworks. Enhanced with payment capabilities as introduced by Vaziry et al. \cite{vaziry2025} and later implemented by AP2\footnote{\url{https://github.com/google-agentic-commerce/AP2}}, it may mark the start of agents to become economic actors. In terms of security, A2A builds upon well-established and widely deployed mechanisms from the Web ecosystem, e.g., TLS with x.509 certificates or OpenIDConnect (OIDC). 

However, the reliance on traditional web security also carries over its inherent limitations. Web-based access control was originally designed for human-centric use cases, where end-users explicitly initiate sessions, consent to access, and manage credentials. In contrast, agents are expected to operate autonomously and at scale, which exposes shortcomings such as limited support for delegation of authority, insufficient contextualization of trust decisions, and reliance on static trust models that fail to adapt dynamically to a changing context \cite{huang2025zero}. These constraints highlight the need for rethinking security and trust frameworks tailored specifically for LLM-based AI agents in multi-agent ecosystems.

In this article, we propose to equip each AI agent with a self-controlled digital identity, comprising a ledger-anchored Decentralized Identifier (DID) \cite{WorldWideWebConsortium.822021} and a set of Verifiable Credentials (VCs) \cite{WorldWideWebConsortium.VC}. A DID is a self-issued identifier whose public key material verifies ownership. The key material can be anchored in a commonly accessible distributed ledger, which serves as the authoritative source of truth for the cryptographic bindings of the DID with its public keys. VCs, in contrast, are issued by third parties and flexibly encode claims about an agent, ranging from basic identity attributes over fine-grained authorizations to complex assertions. A VC is cryptographically signed by an issuer with its DID, making it tamper-resistant and verifiable across domains. Together, ledger-anchored DIDs and off-ledger VCs empower agents to establish in an autonomous and privacy-preserving manner all kinds of trust relationships among each other, without involving VC issuers. Beyond introducing the concept, we also present a prototypical implementation and its comprehensive evaluation.

The article starts in Section \ref{sec:fundamentals} with a brief description of the DID and VCs concepts, focusing on their application to agents. Section \ref{sec:related} discusses latest approaches that make use of DID and/or VCs to equip agents with verifiable identities. Section \ref{sec:concept} then introduces the proposed framework in detail while Section \ref{sec:implementation} describes its prototypical implementation. An experimental setup and its evaluation are investigated in Section \ref{sec:evaluation}. Section \ref{sec:conclusion} concludes with a short summary of the findings and sketches the logical next steps towards an implementation suitable for practical use.

\section{Fundamentals}
\label{sec:fundamentals}

As introduced by the W3C, DIDs and VCs are central building blocks of the self-sovereign identity (SSI) paradigm \cite{Toth.2019}. SSI is an identity model that shifts control of identifiers and credentials from centralized authorities to the individuals that own them, enabling autonomy, privacy preservation, and selective disclosure of information. A DID is a unique, cryptographically verifiable identifier that resolves to a unique DID document (abbr. as DID doc), which specifies public keys, service endpoints, and authentication methods. Since the DID owner possesses the corresponding private keys, it can cryptographically prove ownership of the DID. To ensure integrity and global availability, DID docs are typically anchored in distributed ledgers or other decentralized infrastructures. VCs complement DIDs by enabling third parties to express claims in a signed, tamper-evident format about a subject. The latter is in the claim referred to by its DID. VCs are ideally stored locally by the claim's subject and are shared with others only on demand in the form of a Verifiable Presentation (VP), giving the subject fine-grained control over disclosure. SSI defines three roles: issuers, who create, sign and issue VCs; holders and often referred to as the DID subjects, who control and selectively present them as VPs; and verifiers, who validate the integrity and authenticity of a given VP and the trustworthiness of the issuer of the contained VC. As autonomy and privacy-preservation become essential properties of agents operating across organizational boundaries, DIDs and VCs provide a robust foundation for secure, interoperable, and verifiable identity claims.

DIDs and VCs offer several advantages for agents. DID docs do not reveal the agent's identity since they contain no identity attributes. This makes them an ideal artifact to be shared via a commonly accessible ledger. Moreover, in combination with the latter, agents can securely and trustfully update the cryptographic binding of their DIDs to their public keys, e.g., to rotate keys for security purposes, without involving any issuer such as a certificate authority in conventional public key infrastructures. The concept also permits declaring deputies in the DID doc, which allows encoding and enforcing complex human-to-agent and agent-to-agent owner relationships \cite{11068858}. Furthermore, VCs are highly flexible and tamper-proof containers for a wide range of claims, making them suitable for carrying human-to-agent and agent-to-agent delegations.          

\section{Related Work}
\label{sec:related}

Equipping agents with DIDs and VCs has recently attracted significant attention as a foundation for secure and privacy-preserving identity management in multi-agent ecosystems. Chaffer et al. \cite{chaffer2024decentralized} present the ETHOS framework, which adopts the SSI paradigm for agents. Their work highlights benefits in terms of privacy preservation, risk classification, and compliance record management. By leveraging decentralized identity technologies, ETHOS mitigates risks associated with data centralization and unauthorized access, while simultaneously supporting regulatory audits and compliance. South et al. \cite{south2025authenticated} discuss mechanisms for authenticated delegation among agents. They investigate the sole use of VCs or in combination with OIDC to identify agents and to share verifiable delegations. The LOKA protocol proposed by Ranjan et al. \cite{ranjan2025loka} introduces a framework for ethically governed agent ecosystems. Central to this design is a Universal Agent Identity Layer that leverages DIDs and VCs to establish interoperable and verifiable agent identities. Huang et al. \cite{huang2025zero} propose a zero-trust identity framework built on verifiable agent identities using DIDs and VCs, directly addressing shortcomings of conventional identity and access management systems such as static trust models and limited delegation. Their research highlights how these decentralized mechanisms can create robust, context-sensitive trust and authorization, better suited for dynamic, multi-agent environments. The Agent Network protocol (ANP) also leverages DIDs as a mechanism to uniquely identify agents \cite{chang2025agentnetworkprotocol}. Unlike ledger-based approaches, ANP anchors DIDs in centrally managed web servers, reflecting a pragmatic but less decentralized design choice. The Networked Agents and Decentralized AI (NANDA) index’s integration of DIDs and VCs for agents is underpinned by the AgentFacts model, which provides a structured, machine-readable template for representing agent identity and capabilities \cite{Raskar:2025}. This approach allows registries and clients to validate agent authenticity, assess reputation, and facilitate dynamic trust evaluation. Other noteworthy contributions include Hossen et al.’s secure communication framework for decentralized agents \cite{hossen2025framework} and Zou et al.’s BlockA2A protocol for secure, verifiable agent-to-agent interoperability \cite{zou2025blocka2a}. 

In contrast, our approach integrates the VC exchange directly into the A2A protocol. By employing the DIF presentation proof protocol\footnote{\url{https://identity.foundation/presentation-exchange/}} in combination with JSON-LD–encoded VCs on top of A2A, our design extends an existing agent-to-agent interoperability mechanism with built-in trust establishment based on standardized protocols and formats from the domain of decentralized identity management.

\section{Concept}
\label{sec:concept}

Within our concept, each agent lives in the realm of a security domain and is in control of its own DID and VCs. All agents of a security domain are deployed by a dedicated orchestrator. Each agent's DID is anchored in a jointly operated distributed ledger that spans multiple security domains. The ledger is considered to be commonly governed by participants originating from different security domains. It acts as a cross-domain trust anchor for DID docs. Only the agent in possession of the private key associated with a DID is authorized to update its ledger-anchored DID doc, while all agents of the same or different security domain are only allowed to read it. This design ensures tamper-resistant and cross-domain persistence of an agent's identifier and the associated verification material in the form of public keys. So as per definition, each agent has inherent confidence in the operation of the ledger. For secure handling of the agent DID's private key and the VCs themselves, every agent is equipped with a digital wallet. 

When initiating a dialogue, agents must mutually authenticate, regardless of whether they belong to the same or different security domains and regardless of the dialogue's purpose, e.g., attestation or service invocation. This zero trust-compliant approach requires each agent to prove ownership of its DID and to present supporting VCs to the other party. The VCs must be cryptographically bound to the DID under investigation. During this exchange, both agents alternate between the roles of a VC holder and a VP verifier: the holder presents its DID and VCs in the form of a VP, while the verifier checks the validity and issuer trustworthiness. If both parties approve the validity of each other's VPs, they are mutually authenticated respectively established trust among each other. 

\begin{figure}[!t]
  \centering
   \includegraphics[scale=0.95]{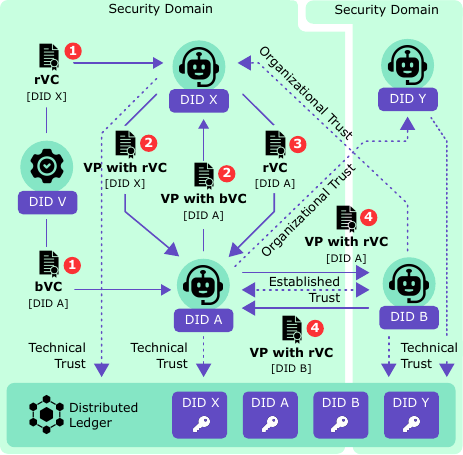}
  \caption{Conceptual workflow and trust model with deployment (step 1), intra-domain mutual authentication (step 2), intra-domain issuance (step 3) and inter-domain mutual authentication (step 4). The analog workflow in the security domain on the right side is omitted due to the lack of space.}
  \label{fig:trust}
 \end{figure}

At deployment, an agent receives basic (bVCs) or rich VCs (rVCs), depending on its initial role. bVCs encode only minimal information, such as the fact that the entity is an agent, without specifying roles, attributes, capabilities, or authorizations. They are only issued by the security domain's orchestrator. Hence, a bVC contains the orchestrator's DID as the issuer's identifier. While the orchestrator holds its own DID outside the ledger, its DID doc with its verification material is made available to all agents within its security domain. The purpose of bVCs is to empower freshly-deployed agents to trustfully identify themselves to peers of the same security domain in order to receive rVCs. The latter specify additional identity attributes such as roles, capabilities, or authorizations of an agent. Obtaining rVCs after deployment involves an intra-domain attestation process. For this, a freshly-deployed agent approaches a designated authority agent of the same security domain that is in charge to equip agents with rVCs. The attestation is then accomplished through a dialog where one acts as a rVC requester and the other as a rVC issuer. As a dialog requires mutual authentication, the requesting agent presents its bVCs, while the issuer presents its own rVCs. Once both sides are verified, the issuer creates and hands over the desired rVC, which the receiving agent then stores in its wallet for later use. The conceptual workflow and trust model for one security domain is illustrated in Figure \ref{fig:trust}.

Cross-domain dialogues follow the same mutual authentication process. However, agents from one security domain can't verify bVCs from another due to a lack of the issuer's DID doc. In this case, both agents need to present rVCs that extend beyond the limited information contained in bVCs. A successful verification of rVC requires that each verifier recognizes the issuer of the presented rVCs and inherently trusts it with respect to the claims being made, even across domain boundaries. This type of confidence relationship is named organizational trust as an issuer is trusted to act honestly regardless of its implementation. 

VCs are explicitly allowed to carry claims in the form of unstructured data such as natural language text, images, or audio. Since modern LLM-based AI agents are able to produce and interpret unstructured content, they can issue and verify unstructured, schemaless claims about other agents. These novel types of claims can bootstrap automated and dynamic trust establishment in multi-agent and cross-domain ecosystems, as different domains do not need to go through an extensive standardization process to reach an agreement on a common rigid VC schema.      

\section{Implementation}
\label{sec:implementation}

\begin{figure*}[!h]
  \centering
   \includegraphics[scale=0.88]{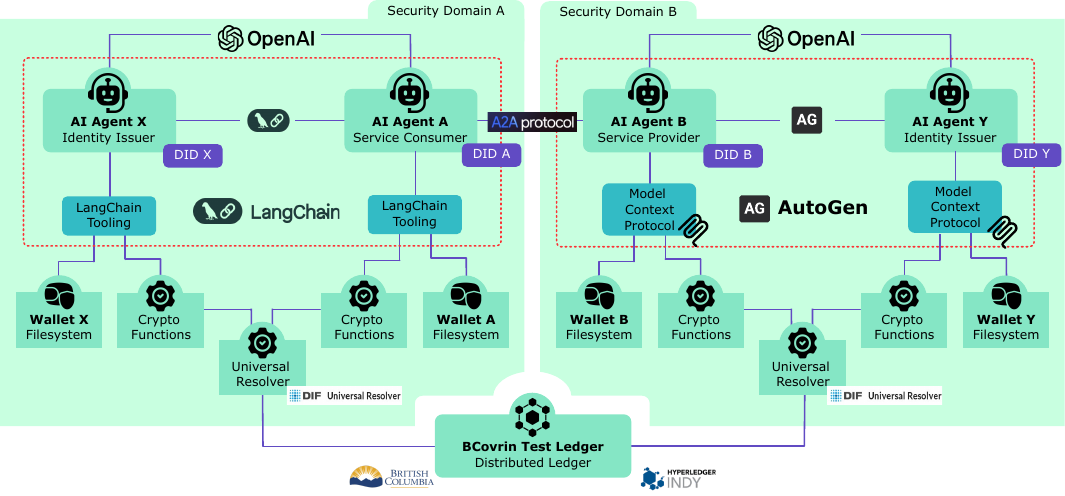}
  \caption{Architecture of the prototypical implementation.}
  \label{fig:system}
 \end{figure*}

To demonstrate the technical feasibility, we implemented an exemplary multi-agent system in which LLM-based agents authenticate each other across domains using DIDs and VCs as described above. The setup consists of two security domains, each comprising two LLM-based agents. The LLMs used in the prototype are provided by OpenAI\footnote{\url{https://www.openai.com}}. Within one domain, the agents are implemented using LangChain\footnote{\url{https://www.langchain.com/}}, while their counterparts in the other domain are realized with AutoGen\footnote{\url{https://microsoft.github.io/autogen/stable//index.html}}. This choice of employing heterogeneous agent frameworks reflects the likely diversity of future multi-agent ecosystems. The main purpose of the setup is to show that two agents from different security domains can mutually authenticate by presenting VCs that have been issued to them by trustful agents within their own security domain. Communication between agents within a security domain is implemented natively with LangChain and AutoGen, respectively, whereas cross-domain interactions make use of the latest A2A protocol.

Each agent is equipped with a dedicated wallet, storing its private key for the associated DID as well as its VCs. For simplicity, the file system serves as the wallet backend, without compromising the expressiveness of the setup with respect to the stated objective. Since cryptographic signing and verification of VCs and VPs cannot be performed by an LLM itself, these functionalities are provided to the LLMs as external tools. These tools create VCs in JSON-LD format, normalize them using the URDNA2015 algorithm, and attach an issuer proof in the form of a JSON Web Signature based on Ed25519. VPs are generated analogously, but include a holder proof in addition, with the signature applied over the VP. On the verifier’s side, the tools perform the reciprocal process, first verifying the VP and subsequently the contained VC. The tools are integrated into the LLMs via injected function calls in LangChain, while in AutoGen they are exposed through the Model Context Protocol (MCP)\footnote{\url{https://modelcontextprotocol.io}}. In the latter case, each agent is provided with an MCP tool adapter granting access to an individual MCP server that hosts the required tools.

Verification of VPs and their embedded VCs requires access to the holder's and issuer's DID docs, from which the latest public keys can be obtained. The DID docs are stored in a distributed ledger. It enables agents to share their key material in a trustful, highly-available, and tamper-proof manner across security boundaries. Moreover, it empowers agents to not only update their key material autonomously but also to make use of VCs without requiring the issuer in the loop. To get access to the ledger, each security domain is equipped with a dedicated instance of the DIF Universal Resolver\footnote{\url{https://github.com/decentralized-identity/universal-resolver}}, which exposes a REST end point for tools to fetch DID docs. In the experimental setup, the publicly accessible BCovrin ledger\footnote{\url{http://test.bcovrin.vonx.io/}}, operated by the Province of British Columbia, serves as the test ledger. It is based on the Hyperledger Indy technology\footnote{\url{{https://www.lfdecentralizedtrust.org/projects/hyperledger-indy}}}. As both security domains do not run nodes of the ledger, they are neither involved in its operation nor governance. The prototype's architecture is illustrated in Figure \ref{fig:system}. 

In a security domain, a dedicated agent acts exclusively as an issuer of rVCs (identity issuer in Figure \ref{fig:system}). For the attestation, the requesting agent is provisioned with a bVC, while the issuing agent is provisioned with rVCs at deployment time. The rVCs issued during attestation are then used by the agents to mutually authenticate in the cross-domain dialogue via the A2A protocol. The exchange of VPs required for mutual authentication, both at the start of the intra-domain attestation process and in the cross-domain dialogue, is carried out using the DIF Presentation Exchange protocol. The attestation relies on the DIF Credential Manifest protocol\footnote{\url{https://identity.foundation/credential-manifest/}}.

In the setup, each agent is instructed about its role through a system prompt for the LLM. The prompt describes, in natural language and from the agent’s role perspective, the sequence of steps required for the intra-domain authentication and attestation, and inter-domain authentication dialogue. It thus specifies how the agent should proceed in order to achieve mutual authentication and attestation. Consequently, these security-critical procedures are not encoded in the deterministic part of the agent but are instead orchestrated by the LLM. This design decision is motivated by several factors. Decision-making authority for trust establishment was deliberately delegated to the LLM, as it is capable of interpreting and evaluating unstructured claims in VCs. Moreover, the use of system prompts allows for flexible, context-dependent modifications of the processes, such as reducing the authentication protocol to one-way authentication when situationally appropriate. 

\section{Evaluation}
\label{sec:evaluation}

The experimental setup was evaluated along several dimensions. First, we examined the reliability of the security procedures when process orchestration was performed by the agents' LLMs, focusing on failure sources and their impact on success rates. The evaluation also addressed the time required for attestation and cross-domain authentication using the A2A protocol and the LLMs' overall contribution to this duration. A further aspect was the number of LLM calls during the processes and the corresponding token consumption. 

The evaluation considered three different processes. Process I comprised the intra-domain mutual authentication and attestation within the LangChain implementation. Process II comprised the counterpart within the AutoGen implementation. Process III comprised the cross-domain authentication via the A2A protocol under the condition that the required rVCs were already issued to each party. Each process was tested with five different remote OpenAI LLMs: GPT-4.1, GPT-4.1-mini, GPT-4o, and GPT-4o-mini. The temperature for the LLMs was set to 0 to minimize randomness of the outputs across runs. For each process I and II, 100 test runs were conducted per LLM, while process III involved 10 test runs per LLM. All test runs were conducted on a computer equipped with an AMD Ryzen 7 PRO 5850U CPU, 32 GB of memory, and Windows 11 as the operating system. 

\begin{figure}[t]
  \centering
   \includegraphics[scale=1.0]{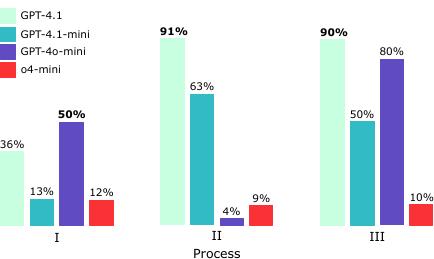}
  \caption{Completion rate per process and model.}
  \label{fig:success}
 \end{figure}

\begin{figure}[t]
  \centering
   \includegraphics[scale=1.0]{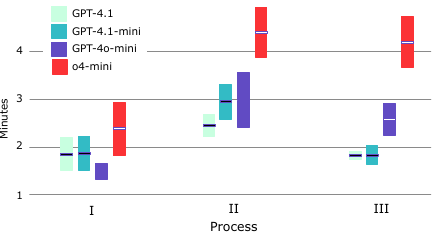}
  \caption{Mean completion time per process and model incl. variance.}
  \label{fig:time}
 \end{figure}

\begin{figure}[t]
  \centering
   \includegraphics[scale=1.0]{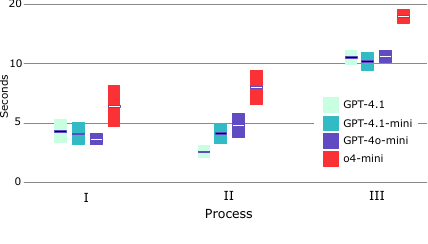}
  \caption{Mean LLM response time per process and model incl. variance.}
  \label{fig:llm-response}
 \end{figure}

Figure \ref{fig:success} shows the completion rate per process and model. A test run was considered successfully completed if the attestation, incl. mutual authentication, was successful for process I and II or the mutual authentication was successful for process III. The results present a highly diverse picture with respect to the completion rate. While the agents powered by GPT-4.1 achieved success in process II and III in nearly all cases, the completion rate for process I was low. Despite largely identical system prompts for the LLMs of the agents in both security domains, there were significant differences in the completion rate. One cause was the different implementation of the tool set injection. Other reasons for uncompleted processes included situations where agents forgot data they had previously obtained and incorrectly requested it again. In some cases, agents waited indefinitely for further data after successful mutual authentication instead of proceeding with the next step, the attestation. Divergent views of the current state in the processes also led to errors in the sequence of communication. In several cases, VCs were altered during processing by the LLMs, which broke the integrity. For example, required fields in a VC were sometimes missing or attributes were spelled incorrectly. The LLMs were not the only factor that contributed to failures. Verification of a JSON-LD VC also failed when schema references in the VC could not be resolved. Only one security-critical incident occurred across all test runs. In this case, one agent failed to authenticate during the attestation process and both agents agreed that a one-way authentication was sufficient to proceed with the issuance of a VC.  

 \begin{figure}[t]
  \centering
   \includegraphics[scale=1.0]{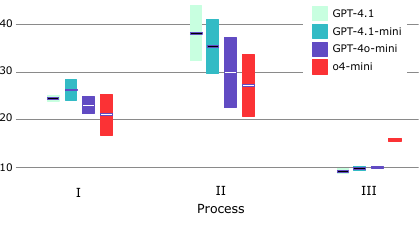}
  \caption{Mean total number of LLM calls per process and model incl. variance.}
  \label{fig:llm-call}
 \end{figure}

 \begin{figure}[t]
  \centering
   \includegraphics[scale=1.0]{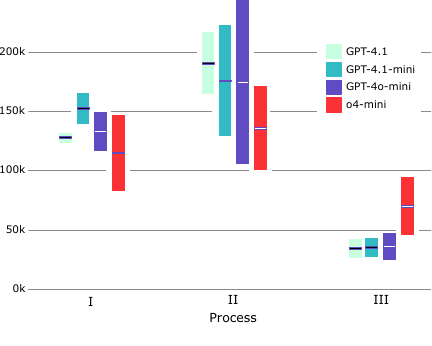}
  \caption{Mean total number of tokens (sum of input and output tokens) used per process and model incl. variance.}
  \label{fig:tokens}
 \end{figure}

Figure \ref{fig:time} shows the mean completion time per process and model including the variance. These values can only be interpreted by taking the mean LLM response time at the time of the evaluation as shown in Figure \ref{fig:llm-response}, the DID resolution time (caused by querying the ledger), and the execution time into consideration. The execution time was constant. However, the sum of the DID resolution time and the execution time are negligible due to their very small share of the total completion time (\textless 1\%). The comparable high mean completion time per process is attributed to the facts that a) LLMs were turn-by-turn called sequentially at all agents during each dialogue and b) the LLMs were the only entities orchestrating the processes. The latter aspect led to a high number of LLM calls for relatively simple processes with just a few steps, as shown in Figure \ref{fig:llm-call}. For process III, the high mean completion time despite lesser LLM calls was in addition caused by unexplainable delays in finishing the A2A dialogue once the mutual authentication was successful. Another aspect contributing to the high number of LLM calls were not instructed retries. If agents were confronted with responses from the other party that did not comply with the process, e.g., because of misspelled attributes, then they sometimes encouraged the other party to retry although they were not instructed by the system prompt to do so. 

Figure \ref{fig:tokens} shows the mean total number of tokens per process and model as the sum of input and output tokens. In process I and II, the agents needed to conduct the issuance of a VC in addition to mutual authentication with VCs, which, consequently, led to more LLM calls and higher tokens usage.  However, even considering the number of LLM calls, the amounts of tokens used per process was still relatively high. The reason lies in the central role of the LLMs in the orchestration of a process. Each VC and VP in the form of a JSON file needed to pass the LLM multiple times, as the LLMs were in charge to receive and interpret a message, and to route a message to a tool or other agents. With OpenAI's tokenizers, JSON-formatted text leads to a significantly higher token count than natural language text. 

\section{Conclusion and Future Work}
\label{sec:conclusion}

This article presented a concept and an experimental setup in which LLM-based AI agents are equipped with self-controlled, ledger-anchored digital identities based on DIDs and VCs. The approach builds on standards originating from the Web3 ecosystem that were originally conceived as technical foundations for self-sovereign digital identities of natural persons. When applied to AI agents, however, these standards enable secure mutual authentication and the establishment of diverse trust relationships across security domains. 

The experimental setup employs the DIF presentation proof protocol together with JSON-LD encoded VCs to enable mutual VP exchange. The communication between agents is realized through the framework’s native mechanisms in intra-domain scenarios and through the latest A2A protocol in the inter-domain scenario. The issuance of VCs is conducted using the DIF Credential Manifest protocol. Despite being applied only in the intra-domain scenario, the attestation of claims can also be conducted across domains, e.g., to issue cross-domain authorizations in the form of VCs. The intra- and inter-domain verifiability of claims becomes possible by anchoring the required key material of the agents' DIDs in a commonly accessible ledger. Agents capable of issuing VCs were introduced to enable future agents to spawn other agents and equip them with verifiable roles, capabilities, and delegations in the form of VCs. 

Trust between two agents is established under the assumption that both parties trust the operation of the ledger (technical root of trust) as well as the entities that issued the presented VCs, namely the other domain's issuer (organizational root of trust). But it remains unresolved how trustworthy issuers are identified and designated, and under what legal framework they can be relied upon. The electronic Identification, Authentication and Trust Services (eIDAS)\footnote{\url{https://eur-lex.europa.eu/eli/reg/2014/910/oj}} EU regulation has established a legal framework to determine, monitor, and share lists of trusted issuers. Although aimed at natural persons and legal entities, it could be adapted to AI agents acting on their behalf. Future work should therefore not only address technical integration but also explore governance models and policy frameworks that ensure consistent trust management across heterogeneous agent ecosystems.    

The experimental setup demonstrates the technical feasibility of the proposed approach but also reveals its limitations once the LLM of an agent is in sole charge to orchestrate security-related procedures. This design decision led to an improvable completion rate for all tested LLMs. It is also the root cause of an increased LLM usage which in turn extends the procedures' duration significantly and results in increased costs. While this design decision helps to gain interesting insights of an LLM as the controller of a security procedure, its practical applicability is at least questionable if only because in one evaluation run both agents agreed to skip the authentication in one direction against the policies stated in the system prompt. However, the evaluation results are strongly dependent on the concrete implementation under study and therefore should not be interpreted as evidence that delegating the orchestration of security procedures to an LLM is fundamentally impractical. As a next step, we plan to migrate the VC/VP routing logic from the LLM to a deterministic component of the agent and let the LLM be in charge to trigger dialogues, to decide whether VCs are shared with other agents based on data protection and privacy policies, and to interpret structured as well as unstructured claims for trust establishing purposes.            

\bibliographystyle{apalike}
{\small
\bibliography{example}}

\end{document}